\documentclass[journal=jpcafh,manuscript=article]{achemso}
\usepackage[version=3]{mhchem} % Formula subscripts using \ce{}
\usepackage{natbib}
\usepackage{notes2bib}

\def\Am3{\AA$^{-3}$}
\def\ph2{{\it p}-H$_2$}
\def\od2{{\it o}-D$_2$}
\def\oh2{{\it o}-H$_2$}
\author{Fabio Mezzacapo}
\affiliation[Max-Planck-Institut]{Max-Planck-Institut f\"ur Quantenoptik, Hans-Kopfermann-Str.1, D-85748, Garching, Germany}
\author{Massimo Boninsegni}
\affiliation[University of Alberta]{Department of Physics, University of Alberta, Edmonton, Alberta, Canada T6G 2G7}
\email{m.boninsegni@ualberta.ca}
\title[]
  {On the possible {``supersolid''} character  \\ of {para}hydrogen clusters}

\abbreviations{IR,NMR,UV}

\begin{document}

\begin{abstract}
We present results of a theoretical study of structural and  superfluid properties of  parahydrogen (\ph2)  clusters comprising 25, 26 and 27 molecules at low temperature. The microscopic model utilized here is based on the Silvera-Goldman pair potential. Numerical results are obtained by means of Quantum Monte Carlo simulations, making use of the continuous-space Worm Algorithm. The clusters are superfluid in the low temperature limit, but display markedly different physical behaviours. For $N$=25 and 27, superfluidity at low temperature arises as clusters melt, i.e., become progressively liquid-like as a result of quantum effects. On the other hand, for $N$ = 26 the cluster remains rigid and solid-like.
We argue that the cluster (\ph2)$_{26}$   can be regarded as a mesoscopic ``supersolid''. This physical picture is supported by results of simulations in which a single \ph2 molecule in the cluster is isotopically substituted.
 \end{abstract}

\section{Introduction}
In a wide range of experimentally relevant, easily accessible thermodynamic conditions, hydrogen molecules can be regarded to an excellent approximation as Bose particles. Their low mass (half that of a helium atom for parahydrogen (\ph2), or equal to that for orthodeuterium (\od2)), and the ensuing significant quantum effects,  led to the expectation that a fluid of such molecules ought to be a ``second superfluid'' (condensed helium being, of course, the one and only known thus far) at sufficiently low temperature.\cite{Ginz72} However, because  the depth of the attractive well of the interaction between two hydrogen molecules is about three times  that between two helium atoms, bulk molecular hydrogen freezes into a non-superfluid crystalline solid  at a temperature $T$ $\sim$ 14 K,  significantly higher than the temperature at which Bose-Einstein Condensation, and the expected resulting superfluid transition, might occur in the fluid. Even though effects of quantum statistics can be detected in the momentum distribution of bulk fluid parahydrogen near freezing,\cite{massimo09} the prospects of observing superfluidity (SF) in the supercooled metastable system appear doubtful, at the present time.\cite{turnbull}

On the other hand,  small molecular hydrogen clusters, with a number of molecules of the order of thirty or less, cannot crystallize in a strict sense, due to their finite size. Consequently, their physical behaviour can remain ``liquid-like'' down to significantly lower $T$ than the bulk, which renders them plausible candidates for the observation of what could be meaningfully named a  superfluid response.  
Over the  past two decades, an intense theoretical and experimental effort has been devoted to the investigation of low $T$ properties of molecular hydrogen \cite{Cep91, Greb00, Fe04, noi06, J06, noi07, Cep07, noi07a, J08, Guard08, noi08, noi09, noi09a, J09, Vil09, Fe09, Cep10, Cep10a, Vil10, McK10}.  
Evidence of SF has been found in \ph2 clusters of 15 molecules,  doped with a single molecular impurity and embedded in helium nanodroplets. \cite{Greb00} In these experiments, the rotational spectrum of the dopant molecule shows evidence of decoupling from the surrounding medium, below $T$=0.19 K. This is in turn interpreted as evidence of superfluidity of the medium itself, i.e., the \ph2 cluster. More recently, these findings have been confirmed in a similar experiment \cite{McK10} performed on doped \ph2 clusters (this time with no helium). Moreover,  new experimental techniques might enable the observation of SF in pristine clusters as well \cite{Fe04, Vil10}. 

From the theoretical standpoint, significant insight has been afforded into the physics of molecular hydrogen clusters, chiefly structure, energetics and superfluid properties, by means of  Quantum Monte Carlo simulations. This methodology has two important advantages: firstly, it furnishes numerical results that for Bose systems are essentially exact, errors only being statistical in character, in practice reducible to very small size  with reasonably modest computational resources. Secondly, it allows for the study of realistic, {\it first principles} microscopic models of the clusters.
Although there is some uncertainty on the choice of the intermolecular potential (we come back to this point later on), the general trend, as observed by simulation, is largely independent of the specific microscopic model utilized. Namely, a robust superfluid signal is predicted for clusters comprising less than $\sim$ 20 molecules, which are liquid-like; their superfluid fraction $\rho_S$ approaches 100\% at $T \simeq$ 1 K. Some clusters in the size range between 23 and 30, display insulating (i.e., non-superfluid), solid-like behaviour at temperatures of the order of a few K, but  {\it melt} at lower $T$ (i.e., become liquid-like) due to quantum effects, concurrently turning superfluid.\cite{noi06, noi07}. SF in these clusters is underlain by long cycles of exchange of identical molecules, involving a number thereof comparable to that of molecules in the cluster. As a result, SF is not confined to any particular spatial region of the cluster, but is essentially uniform throughout it.\cite{noi07a, noi08}.

One of the most intriguing feature of the physics of \ph2 clusters, is the relation between their structural and superfluid properties. Structural changes accompanying  the onset of SF, are particularly evident in clusters which display quantum melting, i.e., go from solid- to liquid-like as $T$ $\to$ 0 K.\cite{noi06, noi07} In particular, the softening of the shells accompanies the appearance of long exchange cycles, not only among molecules in the same shells but also between molecules in different shells. This ``quantum melting'' is seen directly in Monte Carlo simulations, as the system switches back and forth between an insulating solid-like and a superfluid liquid-like phase.

In this paper, we propose that specific \ph2 clusters may possess simultaneously solid and superfluid qualities, and might thus be regarded as ``supersolid'', within the obvious limitations of any such a denomination utilized to characterize a finite system.\bibnote{It is worth stating explicitly that the use of the word ``phase'' itself, albeit common in this field, must necessarily be interpreted loosely in the context of a cluster, for which no well-defined phase exists in the strict thermodynamic sense.} We arrive at this conclusion by studying the physical behaviours of the three clusters  (\ph2)$_{25}$, (\ph2)$_{26}$ and (\ph2)$_{27}$, as they emerge from calculations based on the most realistic and accurate pair interaction models. Specifically, we compare  superfluid fractions and radial density profiles of these clusters as a function of $T$, computed by numerical simulations based on the continuous-space worm-algorithm.\cite{wa, wa1} These three clusters are all superfluid as $T$ $\to$ 0 K;   however,  while the structures of (\ph2)$_{25}$ and (\ph2)$_{27}$ become increasingly liquid-like when $\rho_S$ increases, as clusters undergo quantum melting,  that of  (\ph2)$_{26}$ remains solid-like, essentially temperature-independent in the 0.1 K $-$ 2 K range, in which $\rho_S$ goes from less than 10\% to 100\%. At low $T$, (\ph2)$_{26}$ is  supersolid (i.e., entirely superfluid but also solid-like);  conversely, the two clusters comprising just one fewer or extra molecule are superfluid but liquid-like (i.e., with $\rho_S$ still equal one and a much less pronounced structure). It is important to note that, from the energetic standpoint, the cluster (\ph2)$_{26}$ is not predicted to display greater stability than the other two, at zero temperature.\bibnote{The cluster   (\ph2)$_{26}$ was labeled in some works as ``magic'', i.e., featuring enhanced stability.\cite{Cep07} However, such characterization was based on energy estimates computed at finite temperature ($\sim$ 1 K). More recent calculations\cite{noi09} at lower $T$, yield no evidence of greater stability of (\ph2)$_{26}$ with respect to (\ph2)$_{25}$ and (\ph2)$_{27}$,  in the $T\to 0$ limit}

Support for the intriguing idea of a supersolid cluster comes from results of simulations wherein a single  \ph2 molecule in the cluster is substituted by a \od2 or  \oh2  molecule. The ensuing behaviour is very different  in the three systems considered here.  In the (\ph2)$_{26}$ cluster, the superfluid response is drastically suppressed by the substitution, essentially in equal degrees for a single \od2 or \oh2.  The drop in the value of $\rho_S$ is due to the disappearance of long exchange cycles; in this case, the wave function of the dopant molecule inside the cluster is independent of the mass, i.e., there is no greater tendency of either impurity molecules to reside in specific parts of the cluster. The situation is markedly different when a single substitutional \oh2 (\od2) is injected in clusters of 25 and 27 \ph2 molecules. In both cases, the heavier \od2 molecule is found prevalently in the central part of the cluster, affecting the superfluid fraction of the \ph2 component much more significantly than the \oh2 molecule, which is found to reside in the outer shell.  We interpret this as evidence of greater impurity mobility in the clusters (\ph2)$_{25}$ and (\ph2)$_{27}$, which are liquid-like, whereas, in (\ph2)$_{26}$, foreign molecules of comparable size tend to be bound to reasonably well defined  ``lattice'' positions.

The remainder of this paper is organized as follows: in the next section, we describe the physical model adopted to study either a pristine \ph2 cluster or one doped with a foreign molecule. We then 
present and discuss our results, and finally we outline the conclusions.

\section{Model and Methodology}
We model a pure \ph2 cluster as a collection of $N$ molecules. Each molecule is regarded as a point particle, namely a spinless boson. The quantum mechanical Hamiltonian of the system is:
\begin{equation}\label{ham}
H=-\lambda\sum_i^{N}\nabla^2_i+V(\mathbf{R})
\end {equation}
where $\lambda\equiv \hbar^2/2m$ =12.031 K\AA$^2$, $m$ being the mass of a \ph2 molecule, and $\mathbf{R}\equiv  \mathbf{r}_1,\mathbf{r}_2\cdots\mathbf{r}_N$ is a collective coordinate referring to all $N$ particles in the system.  $V({\bf R})$ is the total potential energy of the system, which in our study is expressed as a superposition of terms only involving pairs of particles:	
\begin{equation}
V({\bf R}) = \sum_{i \ <  \  j} \ v({\bf r}_i,{\bf r}_j)
\end{equation}

The modifications to the above model, if a single molecule is replaced by one of \oh2 (which has the same mass), simply consists of ``labeling'' a particle, i.e., regarding it as distinguishable from the other $N$-1, as the interaction of a \oh2  molecule with \ph2 is very nearly the same as that between two \ph2 molecules. The same applies when  a \od2 molecule is substituted in for a \ph2 one, with the only difference given by the value of $\lambda$, which for this dopant molecule  is half that given above.

In this study, we computed thermodynamic properties at finite temperature of the system described by Eq. (1) by means of Quantum Monte Carlo simulations based on the continuous-space worm algorithm.\cite{wa,wa1} Technical details of our calculations are the same as in other works.\cite{wa,wa1,noi06,crb}. Because hydrogen molecules in their ground state are very nearly spherical, it is usually considered a reasonable approximation to regard the potential $v({\bf r},{\bf r^\prime})$ as central, i.e., $v({\bf r},{\bf r^\prime}) \equiv v(|{\bf r}-{\bf r^\prime}|)$.
All of the results presented here were obtained using for $v(r)$ the Silvera-Goldman (SG) pair potential \cite{SG}, the most commonly employed in all previous investigations.\cite{Cep91, noi06, J06, noi07, noi07a}.  Recent Quantum Monte Carlo simulations have shown that the zero-temperature equation of state computed with the SG potential is in reasonable agreement with  experimental results. \cite{op, op1} However, there exist other choices as well, such as the Buck\cite{buck} or the Lennard-Jones (LJ) potential, which have also been used to model these systems.   

The issue of the inter-molecular potential is important enough that a comment is in order. 
The main features of all  these potentials are {\it a}) a ``hard core'' repulsion for inter-molecular distances less than $\sim $ 0.2 nm {\it b}) an attractive potential well, of depth $\sim$ 35 K, centred at approximately 0.3 nm {\it c}) an attractive tail, arising from interaction between mutually induced molecular electric dipoles, decaying as 1/$r^6$ at long distances.
A molecular hydrogen cluster is at the border between a classical and a quantum object; clearly, therefore, small energy differences can (and do) result in large quantitative differences in the estimates of relevant physical properties. This seems to be especially the case with the Lennard-Jones potential. Its less attractive character, compared to the more accurate Silvera-Goldman and Buck potentials, gives rise to floppier and more liquid-like clusters, whose superfluid fraction is typically higher than that obtained with the other two potentials. Furthermore, despite the interaction-independent general trend of $\rho_S$ as a function of cluster size, important {\it qualitative} differences arise for specific clusters, if the LJ potential is used. For example, the cluster (\ph2)$_{26}$ is predicted to undergo quantum melting \cite{noi09a} when the LJ interaction is adopted, in stark contrast with what is found in this work, based on the SG potential. 
At the other extreme there is the Buck potential, which is significantly more attractive than the SG. Clusters are predicted to be more rigid, and those displaying quantum melting remain solid-like down to temperatures significantly lower than those at which SG clusters turn liquid-like. 

To some extent, therefore, the more quantitative aspects of our results do reflect the particular choice of interaction made here, namely the SG potential. However, it is important to restate that, while details of the results may differ, none of the physical effects discussed here (or in our previous works) are a hallmark of the SG interaction -- in other words, they are expected to arise even with the LJ or Buck potentials. Quantum melting, for example, is observed with all of the three potentials mentioned above, albeit at different temperatures and/or cluster sizes.

\section{Results}

We show in Figure 1 the superfluid fraction $\rho_S(T)$ of clusters comprising 25, 26 and 27 \ph2 molecules. The calculation of this quantity makes use of the well-known ``area'' estimator, originally introduced by Sindzingre, Klein and Ceperley.\cite{Cep89}  As shown in the figure, all three clusters are fully superfluid at the lowest $T$ considered here. However  the superfluid fraction saturates to 100\% rather differently, both quantitatively as well as qualitatively,   in the low $T$ limit. In particular  $\rho_S$ $\sim$ 1 at $T$ = 0.5 K for $N$=25, and at  lower $T$ (i.e., 0.25 and 0.125 K respectively) for $N$=27 and $N$=26, illustrating how a difference of just one molecule can strongly affect quantitatively the superfluid behaviour.

 \begin{figure}[h]
\centerline{\includegraphics[scale=0.75]{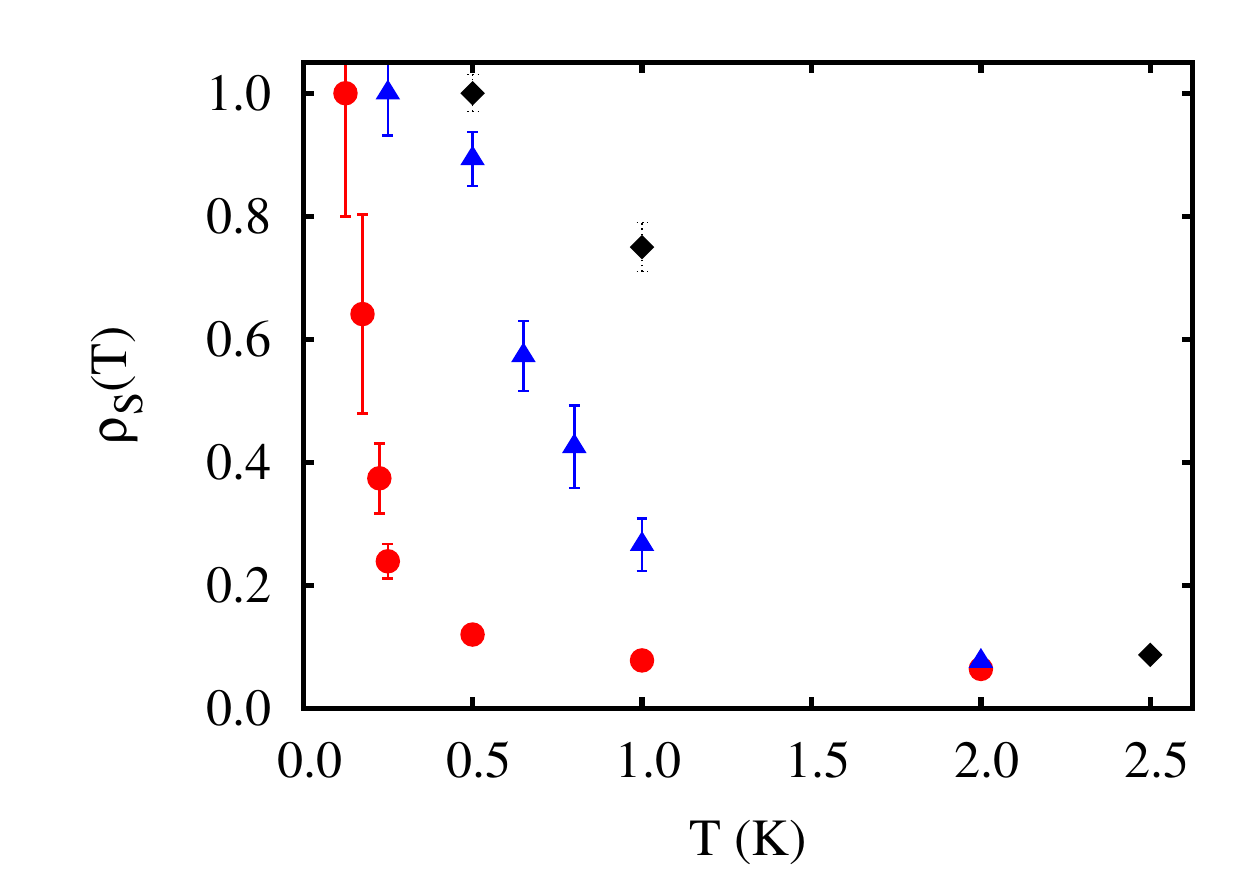}}
\caption{(color online) Temperature dependent superfluid fraction for clusters comprising 25 (diamonds), 26 (circles) and 27 (triangles) \ph2 molecules. Error bars, where not shown, are of the order (or smaller) than the symbol size.}
\label{sup}
\end{figure}

Figure 2 shows values of the superfluid fraction and the potential energy per molecule ($V$) recorded during a typical Monte Carlo simulation for the cluster (\ph2)$_{25}$ at $T$=1.125 K (left panel). Specifically we show consecutive block (each consisting of 500 sweeps\cite{noi07}) averages of $\rho_S$ and $V$. Although noticeable fluctuations affect the values of $\rho_S$ and $V$, as in any Monte Carlo simulation, two different regimes are clearly identifiable, namely one in which the average value of $\rho_S$ is close to 1 and concomitantly the average $V$ is $\sim$ $-59$ K, the other characterized by lower values of both quantities ($\rho_S$ is close to zero, whereas $V$ $\sim$ $-64$ K). In the first regime, the cluster is superfluid and liquid-like,  in the second non-superfluid and solid-like. The kinetic energy per molecule displays similar jumps as the potential energy, anti-correlated so that the total energy remains the same in the two regimes, within statistical fluctuations. 

\begin{figure}[ ]
{\includegraphics[scale=0.625]{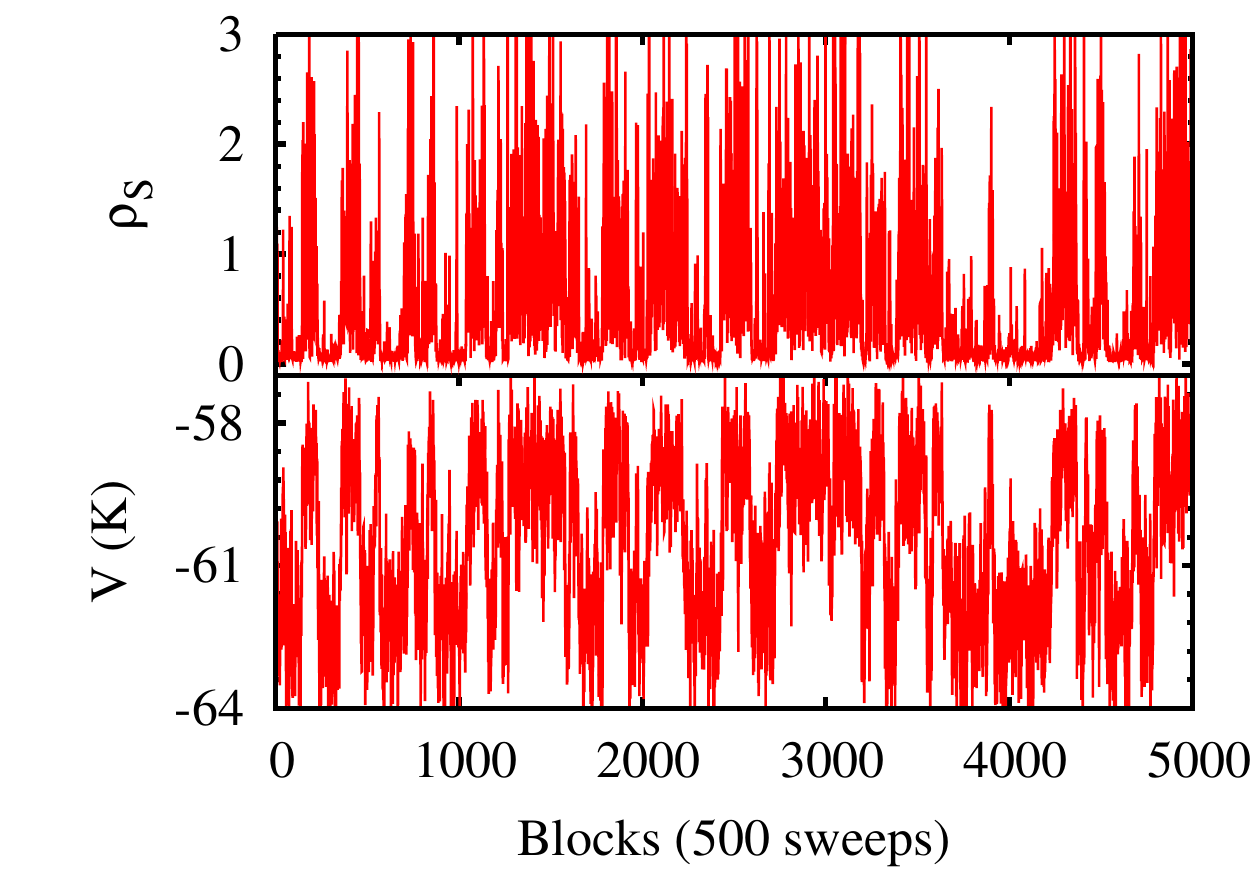}}
{\includegraphics[scale=0.625]{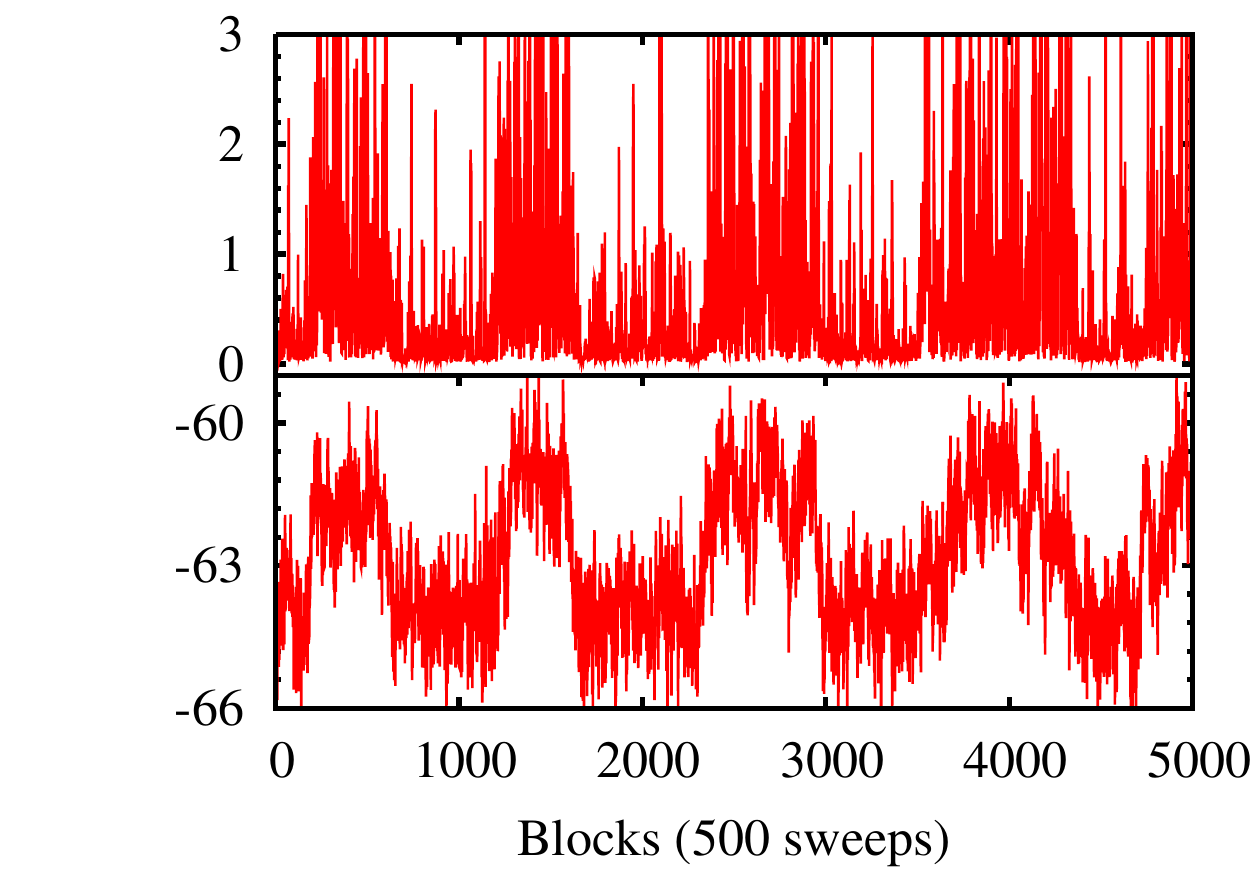}}
\caption{(color online) Left: Potential energy per molecule (lower panel) and superfluid fraction (upper panel) observed during a typical Monte Carlo run for the cluster (\ph2)$_{25}$ at $T$=1.125 K. Two coexisting phases are clearly recognizable since the averages of $V$ and $\rho_S$ switch simultaneously between high (liquid-like and superfluid phase) and low values (solid-like and insulating phase). Right: Same observables for the cluster (\ph2)$_{27}$ at $T$=0.65 K.}
\label{qm25}
\end{figure}

We associate to these two regimes to distinct ``phases'' of the cluster.  On decreasing the temperature, the liquid-like phase becomes progressively dominant, i.e., the cluster ``quantum melts''. The same phase coexistence, 
signaling the onset of quantum melting, is observed for the (\ph2)$_{27}$ cluster (see Figure 2, right panel). On the other hand, for the cluster  (\ph2)$_{26}$, at no value of the temperature in the range considered here, wherein its superfluid fraction  goes from zero to essentially 1, the same evidence that the system switches between two phases shown in Figure 2  has been found. In other words, no evidence of quantum melting is observed for this cluster.

 In order to gain insight into the subtle connection between superfluid and structural properties of these clusters, let us consider cluster radial density profiles, computed with respect to the centre of mass of the system. At $T$ = 2.5 K, the  cluster (\ph2)$_{25}$ displays a structure consisting of two shells, signalled  by peaks at $r\sim$ 2 \AA$ $ and $r\sim$ 5 \AA$ $  (dashed line in Figure 3, left panel). As one can see, there is no clear demarcation between the two, consistently with a relatively ``floppy" structure. On decreasing the temperature, however, the system loses some of its rigidity; at $T$ = 0.5 K (solid line in the same figure) the  first peak is less pronounced, due to the enhanced delocalization of the \ph2 molecules, and concurrently $\rho_S$ increases from  less than 10\% to 100\%. 
The same physical picture holds for (\ph2)$_{27}$ (see Figure 3, right panel). The structural changes occurring in the cluster (\ph2)$_{25}$ and (\ph2)$_{27}$, although quantitatively not dramatic, provide further, qualitative  support to the idea of quantum melting, for which the strongest evidence is given by data shown in Figure 2. In these clusters, structural transformations and SF are strictly related. Specifically both (\ph2)$_{25}$ and (\ph2)$_{27}$ are solid-like and non-superfluid at high $T$, increasingly liquid-like and superfluid as $T$ decreases.

 \begin{figure}[h]
{\includegraphics[scale=0.65]{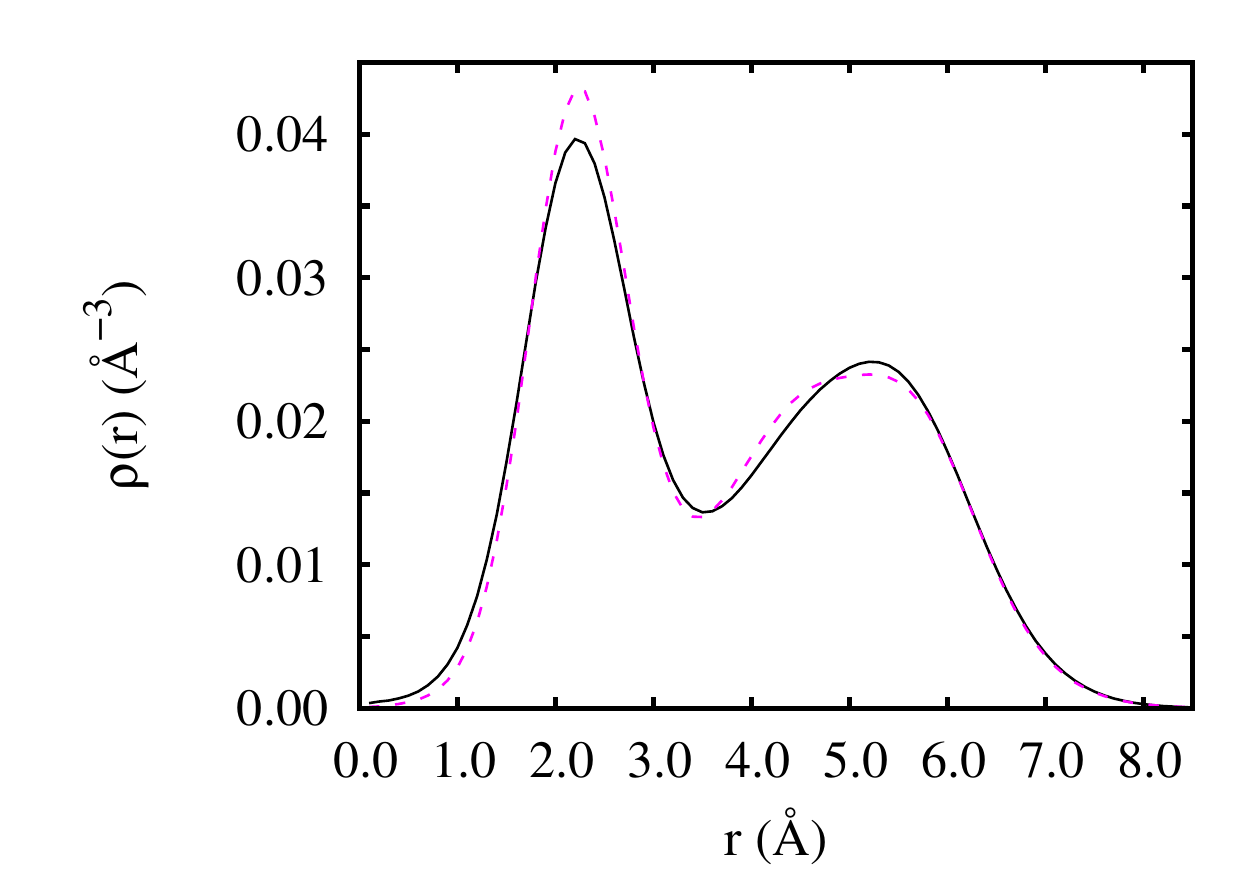}}
{\includegraphics[scale=0.65]{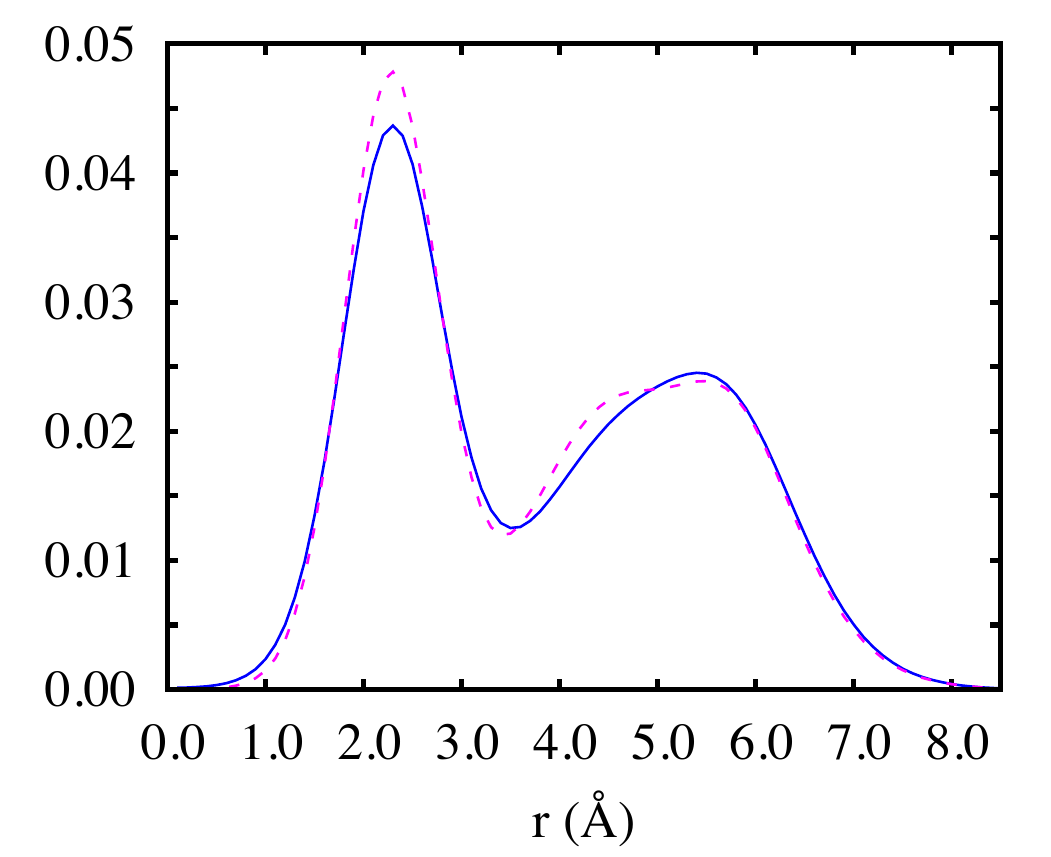}}
\caption{(color online) Left: Radial density profiles for the cluster (\ph2)$_{25}$ at $T$=0.5 K (solid line) and $T$=2.5 K (dashed line). Right: Radial density profiles for the cluster (\ph2)$_{27}$ at $T$=0.25 K (solid line) and $T$=2 K (dashed line).}
\label{r25}
\end{figure}

The situation is remarkably different for the cluster  (\ph2)$_{26}$. Its superfluid fraction increases from $\sim$ 0.06 (at $T$= 2 K) to $\sim$ 1  ( at $T$ = 0.125 K); however, at no temperature in this range is coexistence of two phases observed, of the type shown in Figure 2. Correspondingly,  the radial density profile (shown in Figure 4) remains essentially unchanged.\bibnote{In a previous work by us,\cite{noi08} the cluster (\ph2)$_{26}$ was predicted to undergo quantum melting at low $T$ on the basis of density profile computed by neglecting the motion of the centre of mass of the cluster in imaginary time. While this does not cause major changes in the results at $T$ above $\sim$ 1 K, failing to take it into account at lower temperature has the result of smearing the radial density profiles, which we initially incorrectly interpreted as evidence of quantum melting. We are indebted to D. Ceperley for pointing this out to us.} The superfluid character of the cluster which emerges at low $T$ is independent of  the structure, which is markedly solid-like. In fact, comparing the low temperature density profiles of (\ph2)$_{25}$ and (\ph2)$_{27}$  to that of (\ph2)$_{26}$, it is evident how the latter features a significantly higher first shell peak, as well as greater structure in the second shell. The cluster  (\ph2)$_{26}$ at low $T$ is therefore superfluid and solid-like, properties which coexist by definition in a supersolid. It is interesting to note that the suggestion that a \ph2 clusters might display a physical behaviour warranting such denomination, was made in the pioneering study by Sindzingre, Ceperley and Klein,\cite{Cep91} with reference to (\ph2)$_{13}$, now  believed to be liquid-like.

A possible alternate explanation for the large superfluid response of a cluster,  is that the system may simply behave 
as a rigid rotator; as soon as the temperature is lowered below the characteristic separation between the first two 
energy levels of a spherical rotator with the same classical moment of inertia, the cluster may no longer absorb energy by making transitions to states of higher orbital angular momentum, and would consequently be able to rotate freely, i.e., act as a ``superfluid''. 
The computation of the moment of inertia $I$ carried out in this work, yields a ``transition temperature'' 
$\hbar^2/I \sim 0.05$ K for (\ph2)$_{26}$, which is in the right ballpark, albeit lower than what observed here. However, cycles of exchange including any number $M$ of molecules between 1 and 26, are found\cite{noi08} to take place in the low temperature limit, with comparable frequencies. This seems to rule out the scenario of the rigid rotations, whereby cycles with $M\sim N$ would occur much more frequently, all others being suppressed.

\begin{figure}[h]
\centerline{\includegraphics[scale=0.75]{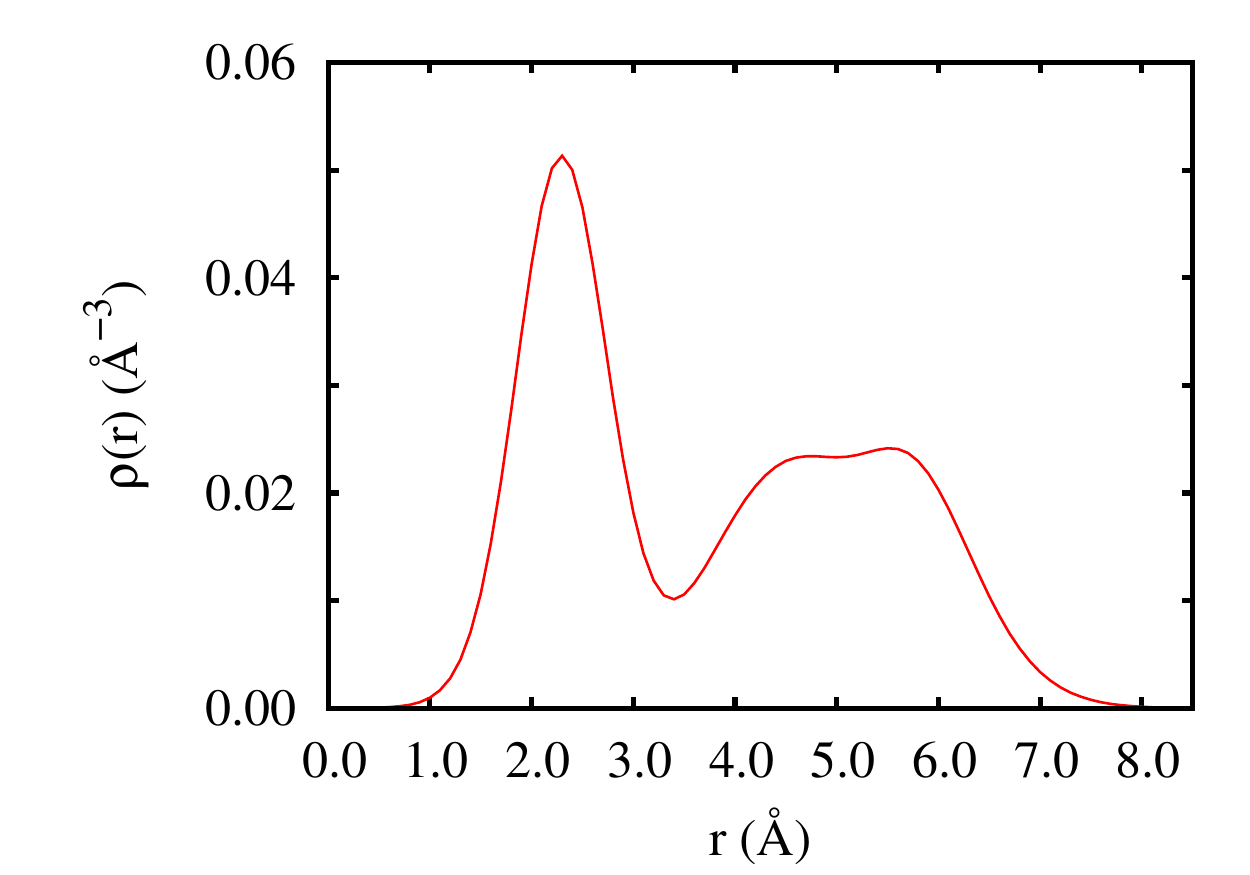}}
\caption{(color online) Radial density profiles for the cluster (\ph2)$_{26}$ at $T$=0.125 K. The structure of this cluster does not show significant changes with the temperature.}
\label{r26}
\end{figure}

In order to characterize in greater depth the  physics of (\ph2)$_{26}$, chiefly its different nature with respect to  (\ph2)$_{25}$ and (\ph2)$_{27}$,    we also studied the effect on the \ph2 superfluid response  of a single substitutional molecule. Specifically, we investigated clusters of size $N$=25, 26 and 27 in which a single \ph2 molecule is replaced by one of   \od2 or  \oh2 (orthohydrogen). As a result of the substitution of  a \ph2 molecule with one of \od2 (two times heavier), the \ph2 superfluid response in (\ph2)$_{27}$ drops by almost 60\% at  $T$=0.25 K; conversely, it remains essentially unchanged when the dopant molecule is one of \oh2, whose mass is the same of a \ph2 molecule. 
 \begin{figure}[!]
\centerline{\includegraphics[scale=0.7]{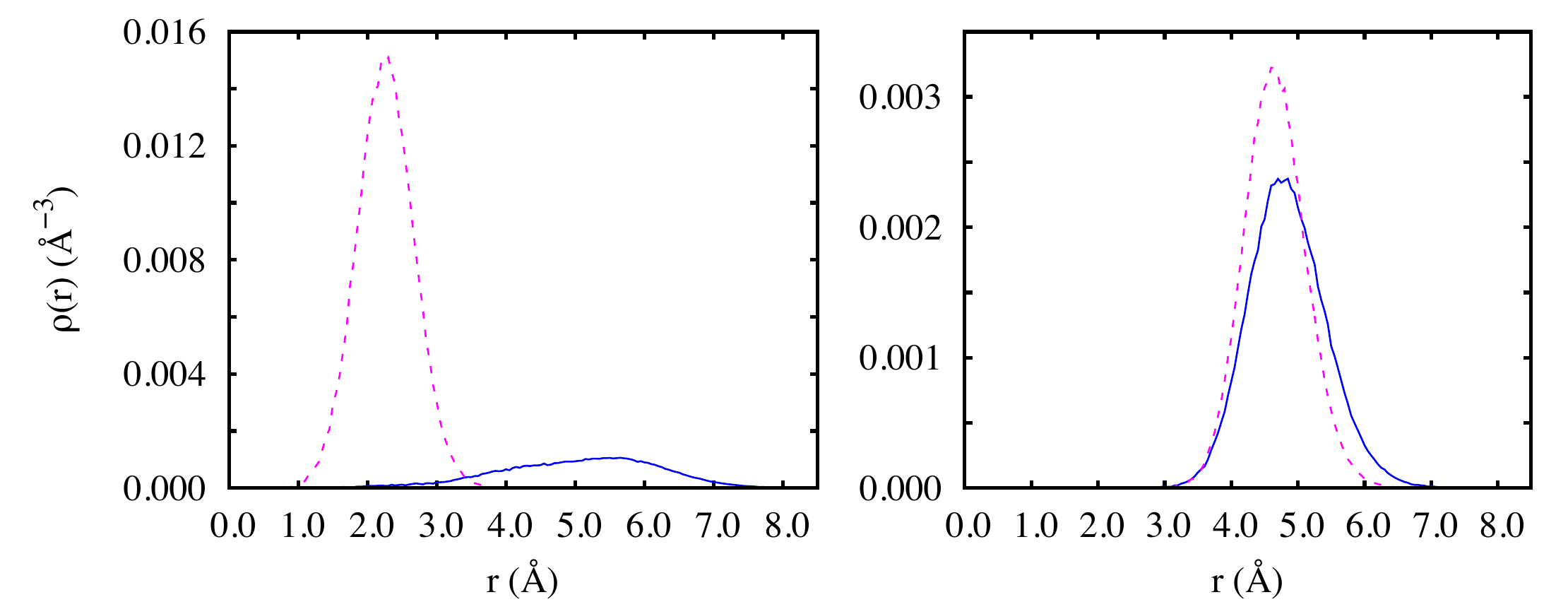}}
\caption{(color online) Density profiles of dopant impurity, computed with respect to the geometrical center of the system, for the doped cluster  (\ph2)$_{26}-$(\od2)$_1$ (dashed line) and  (\ph2)$_{26}-$(\oh2)$_1$ (solid line) at $T$=0.25 K (left), and for  (\ph2)$_{25}-$(\od2)$_1$ (dashed line) and  (\ph2)$_{25}-$(\oh2)$_1$ (solid line) at $T$=0.125 K (right).  }
\label{rad26+1}
\end{figure}

A clear difference is observed in the density profiles of the two dopant molecules, computed with respect to the cluster geometrical centre (left panel of Figure 5). The heavier \od2 is found in the inner part of the cluster,  while the lighter dopant resides prevalently in the outer shell. The mass difference of the dopant, which in both cases interacts with the \ph2 molecules in the same way as a \ph2 molecule,  therefore plays a crucial role and appears responsible for the different superfluid response  of the \ph2 component in the two clusters.
The same physical result is found when a dopant molecule is substituted in the superfluid  (\ph2)$_{25}$ cluster. 
In both cases, the heavier molecule sits in the inner shell, ostensibly because of its greater mass, which makes its localization inside a smaller region energetically less unfavourable than for a lighter molecule.

This is also shown by the three-dimensional representation of the doped clusters with $27$ molecules (Figure 6). Even though the information provided by this kind of figure is mostly qualitative,\cite{sav} structural differences between the two clusters are evident. Specifically, (\ph2)$_{26}-$(\oh2)$_1$ (right) is liquid-like, i.e., molecules are delocalized and the cluster featureless. In this case,  the impurity resides in the outer shell. By contrast,  (\ph2)$_{26}-$(\od2)$_1$ is  solid-like;  molecules are highly localized, even though exchanges still occur, and the impurity in this case is confined to the inner shell. The solid-like structure of (\ph2)$_{26}-$(\od2)$_1$ is consistent with the substantial drop in the superfluid fraction with respect to the pristine cluster. The heavier dopant molecule in the inner shell is much more localized  than the lighter one in the outer shell. 
 In particular, one may infer from the density profiles shown in the left panel of Figure 5 that the \od2 is confined within a shell of thickness $\sim$ 1 \AA, whereas that for the \oh2 is approximately twice as wide. 

The right panel of Figure 5 shows the density profiles of the same impurities, substituted into a (\ph2)$_{26}$ cluster, at low temperature ($T$=0.125 K), at which a pristine (\ph2)$_{26}$ cluster is entirely superfluid. The contrast with respect to the clusters with one extra (less) \ph2 molecule is striking. In this case, there is little difference between the two profiles, i.e., both  impurities reside in the outer shell, at the same distance from the centre and in a spatial region of very similar size ($\sim$ 1 \AA), the heavier \od2 molecule being only slightly more confined. In this case, the   drop of the \ph2 superfluid fraction resulting from the substitution is the same for \oh2 and \od2, within statistical uncertainties,  from 100\% to approximately 20\%. Moreover, lengthy computer simulations yield evidence that diffusive processes are extremely slow in the cluster with 26 molecules.
 \begin{figure}[!]
\centerline{\includegraphics[scale=0.7]{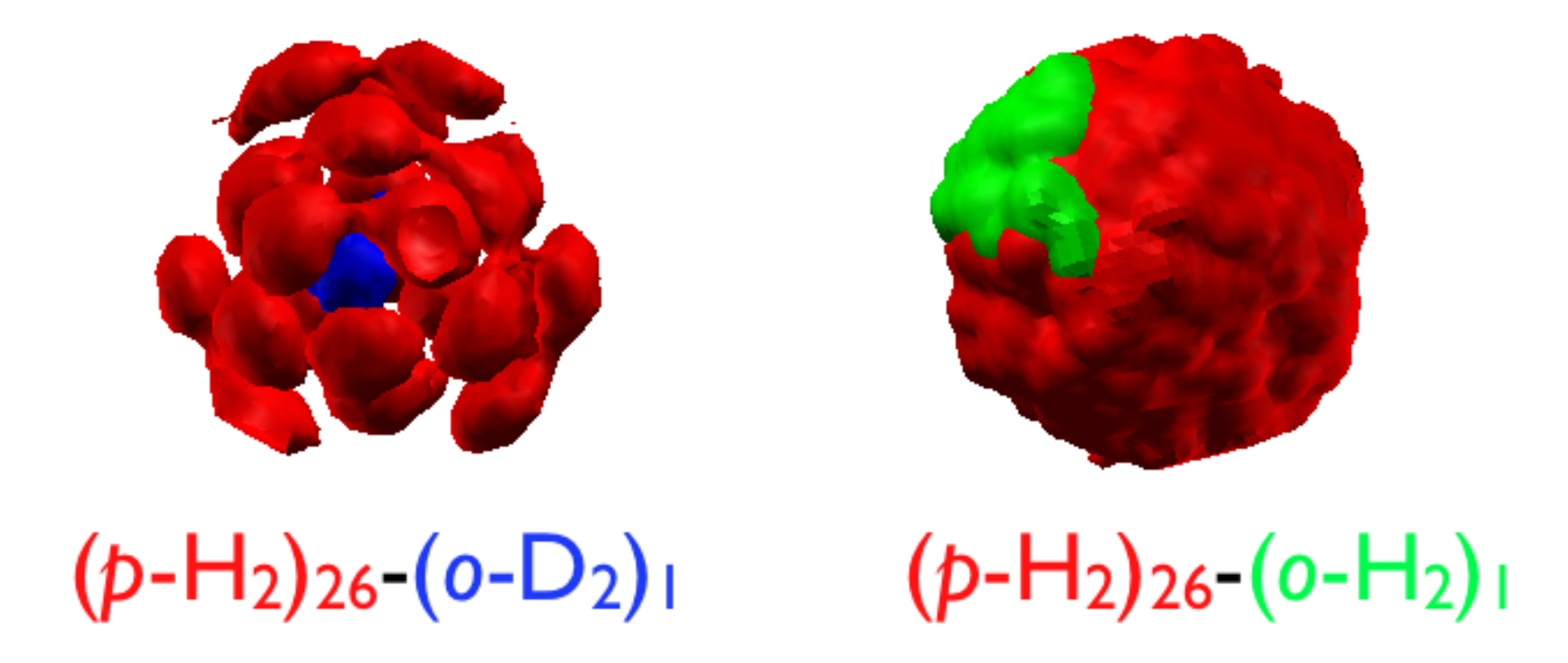}}
\caption{(color online) Three-dimensional representation of the clusters  (\ph2)$_{26}-$(\od2)$_1$ (left)  and  (\ph2)$_{26}-$(\oh2)$_1$ (right)  at $T$=0.25 K. }
\label{pallocchi}
\end{figure}

These findings can be interpreted as follows: \\
{\it a}) the lighter (\oh2) impurity enjoys much greater mobility in a cluster with 27 (or, 25) molecules, than in that with 26. This is in turn consistent with a picture in which the shells of clusters with 25 and 27 molecules are liquid-like and superfluid, allowing for the essentially free motion of an impurity of the same mass. On the other hand, a heavier dopant occupies the inner shell, where it has a more disruptive effect on exchanges of identical molecules, thereby opposing quantum melting (largely underlain by exchanges) and  favouring energetically a more solid-like cluster structure.
\\
{\it b}) the structure of (\ph2)$_{26}$ is one in which molecules occupy preferentially specific ``lattice'' sites. Such localization is consistent with the observed, sluggish (Monte Carlo) dynamics, and not incompatible with quantum-mechanical exchanges of identical particles, which, as is well known, underlie superfluidity. Indeed, in a hypothetical {\it supersolid} phase both aspects are supposed to co-exist. One ought keep in mind that an important distinction between a superfluid and a supersolid, is that the latter does {\it not} allow an impurity to move through it without dissipation.

The frequency of occurrence of exchange cycles comprising $M$ \ph2 molecules in the doped clusters (\ph2)$_{26}$-(\oh2)$_1$ and  (\ph2)$_{26}$-(\od2)$_1$  at $T$ = 0.25 K is shown in Figure 7 (left panel). 
Long exchanges, involving  almost all the \ph2 molecules in the cluster, are largely suppressed in the latter case, remaining instead frequent when the dopant molecule is one of \oh2. These long exchanges underlie the robust superfluid behaviour of  the \ph2 component in the cluster doped with \oh2. Conversely, the absence of such long exchanges is at the root of the depression of the superfluid response in the cluster doped with a \od2. 
 \begin{figure}[t]
 {\includegraphics[scale=0.65]{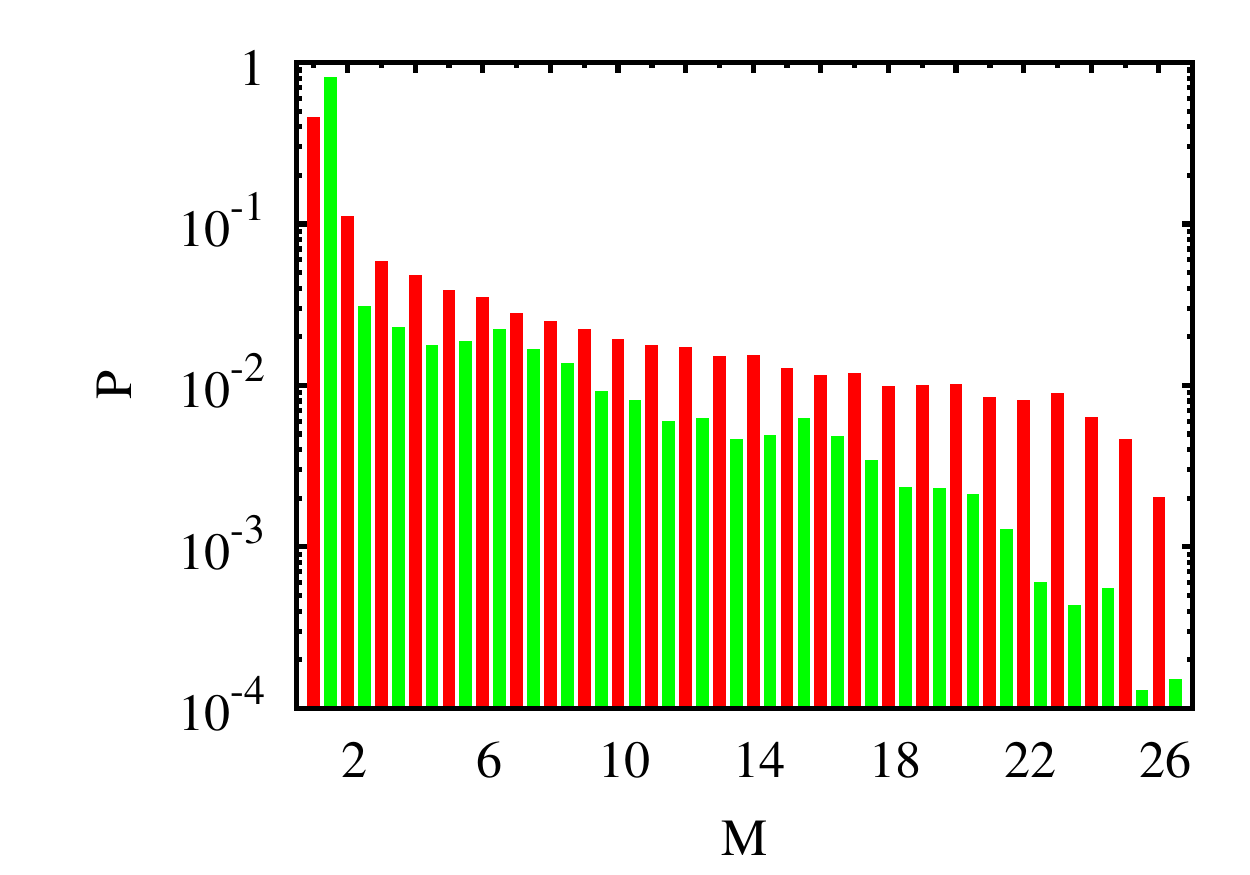}}
{\includegraphics[scale=0.65]{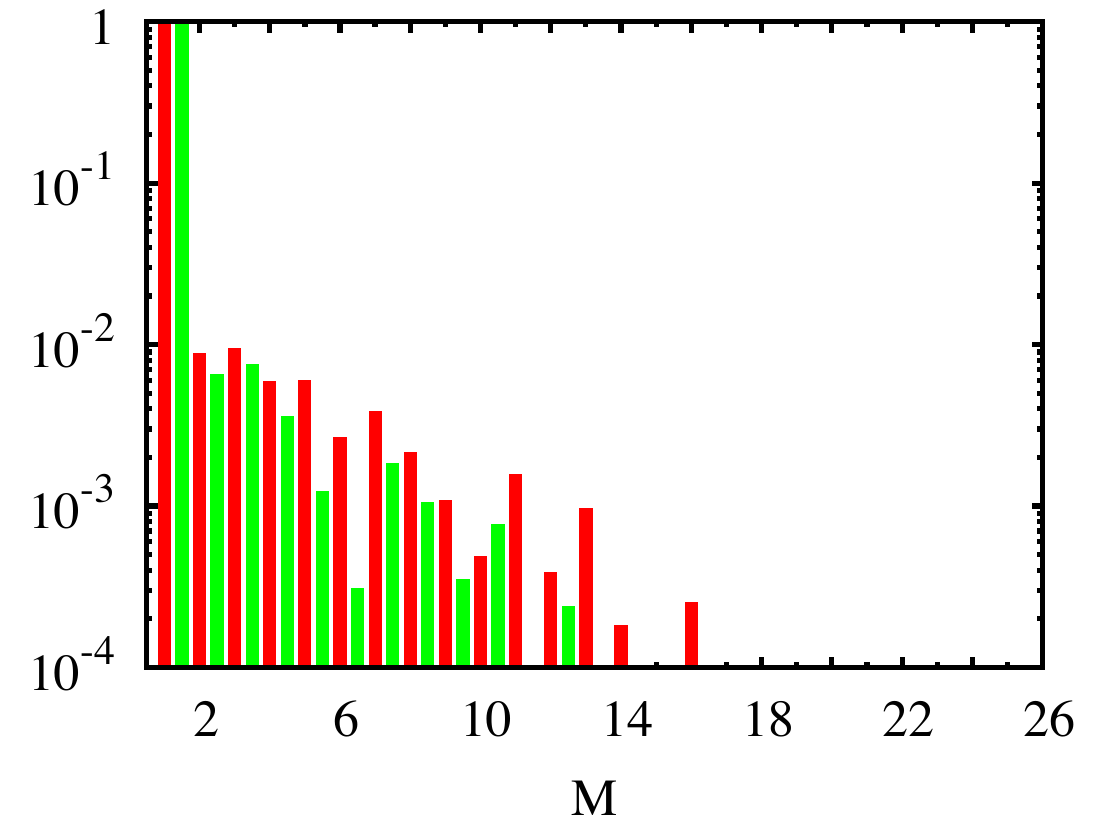}}
\caption{(color online) Frequency of exchange cycles comprising $M$ \ph2 molecules for the doped cluster (\ph2)$_{26}$-(\od2)$_1$ (lighter color) and (\ph2)$_{26}$-(\oh2)$_1$ (darker color) at $T$=0.25 K (left), and for the doped cluster (\ph2)$_{25}$-(\od2)$_1$ (lighter color) and (\ph2)$_{25}$-(\oh2)$_1$ (darker color) at $T$=0.125 K (right).  }
\label{cy26+1}
\end{figure}

The suppression of long exchanges is much more significant in the case of the heavier impurity, which resides in the central shell, than for the lighter one which is in the outer part of the cluster (see Figure 6). The impurity sitting at the centre of the cluster ostensibly has a more disruptive effect over exchanges involving molecules in different shells.  This underscores the importance of such exchanges, in order for the superfluid signal to be large, even though most molecules reside in the outer shell, in a cluster of such a small size. 

On the other hand, in the case of a doped (\ph2)$_{26}$ cluster, the suppression of the superfluid response is essentially identical with both dopant molecules, suggesting that in both cases the dopant simply occupies one of the lattice positions, and its disruptive effects on exchanges of the other, identical particles, is quantitatively similar. This is confirmed by the data on the statistics of exchange cycles  in the doped clusters (\ph2)$_{25}$-(\oh2)$_1$ and  (\ph2)$_{25}$-(\od2)$_1$  at $T$ = 0.125 K, shown in Figure 7 (right panel). {\it Long} exchanges are {\it drastically} suppressed, in fact essentially non-existent in the case of either impurity. The fact that exchanges of intermediate length occur with different frequencies in the two cases, as shown in the right panel of Figure 7,  is less relevant, as these exchanges are not the ones that underlie a large superfluid response.

\section{Conclusions}
Low temperature Quantum Monte Carlo simulations have been carried out to investigate the relation between structural and superfluid properties of \ph2 clusters of size 25 $\le$ $N$ $\le$ 27, modelled using the Silvera-Goldman pair potential. The results presented here show that the (\ph2)$_{26}$ cluster has a qualitatively different behaviour from the other two, displaying simultaneously superfluid and solid-like properties. By contrast, the other two clusters melt at low temperature due to quantum effects. This conclusion is established by studying the evolution of radial density profiles with temperature, as well as by comparing results obtained with substitutional orthodeuterium and orthohydrogen molecules.

As mentioned above, the intermolecular potential adopted in any such calculation can have a significant effect on the quantitative details of the results. In particular the physics observed here for  (\ph2)$_{26}$ is different from that which arise if the Lennard-Jones potential is used, in which case quantum melting is observed.\cite{noi09a} On the other hand, calculations based on the Buck pair potential at a temperature of 0.0625 K yield results consistent with those obtained with the Silvera-Goldman pair potential, i.e., with the predicted supersolid character of  (\ph2)$_{26}$. As the Silvera-Goldman and Buck potentials are believed to afford greater accuracy than the Lennard-Jones potential, in the description of the condensed phase of molecular hydrogen, it is our view that the predictions at which we have arrived here are robust, at least to the extent to which a spherical potential can be regarded as a reliable approximation for these systems. 

Naturally, it is quite possible that similar physics could be observed for greater clusters as well, possibly at lower temperature. However, the results presented here show that (\ph2)$_{26}$ may be the best candidate for the observation of  some of the physics characterizing a ``supersolid'' phase, albeit in a finite system of small size.
The very different physical behaviours predicted here for clusters of 27 (or 25) and 26 molecules should give rise to correspondingly different phenomenology, for example in experiments probing the rotational spectra of linear molecules embedded in pristine (\ph2) clusters. 

\begin{acknowledgement}
This work has been supported by the DFG (FOR 635) and the EU project QUEVADIS, and the Canadian NSERC through the grant G121210893.
\end{acknowledgement}

\bibliography{achemso-demo}

\begin{mcitethebibliography}{36}
\bibitem[Not()]{Note-1}
It is worth stating explicitly that the use of the word ``phase'' itself,
  albeit common in this field, must necessarily be interpreted loosely in the
  context of a cluster, for which no well-defined phase exists in the strict
  thermodynamic sense.\relax
\mciteBstWouldAddEndPunctfalse
\mciteSetBstMidEndSepPunct{\mcitedefaultmidpunct}
{}{\mcitedefaultseppunct}\relax
\EndOfBibitem
\bibitem[Not()]{Note-2}
The cluster (\ph2)$_{26}$ was labeled in some works as ``magic'', i.e.,
  featuring enhanced stability.\cite{Cep07} However, such characterization was
  based on energy estimates computed at finite temperature ($\sim$ 1 K). More
  recent calculations\cite{noi09} at lower $T$, yield no evidence of greater
  stability of (\ph2)$_{26}$ with respect to (\ph2)$_{25}$ and (\ph2)$_{27}$,
  in the $T\to 0$ limit\relax
\mciteBstWouldAddEndPuncttrue
\mciteSetBstMidEndSepPunct{\mcitedefaultmidpunct}
{\mcitedefaultendpunct}{\mcitedefaultseppunct}\relax
\EndOfBibitem
\bibitem[Not()]{Note-3}
In a previous work by us,\cite{noi08} the cluster (\ph2)$_{26}$ was predicted
  to undergo quantum melting at low $T$ on the basis of density profile
  computed by neglecting the motion of the centre of mass of the cluster in
  imaginary time. While this does not cause major changes in the results at $T$
  above $\sim$ 1 K, failing to take it into account at lower temperature has
  the result of smearing the radial density profiles, which we initially
  incorrectly interpreted as evidence of quantum melting. We are indebted to D.
  Ceperley for pointing this out to us.\relax
\mciteBstWouldAddEndPunctfalse
\mciteSetBstMidEndSepPunct{\mcitedefaultmidpunct}
{}{\mcitedefaultseppunct}\relax
\EndOfBibitem
\end{mcitethebibliography}

\end{document}